\documentclass[a4paper,12pt]{article}
\usepackage{graphicx}

\title{
Taming a Chaotic Dripping Faucet via a Global Bifurcation
}
\author{
Ken Kiyono\thanks{E-mail: k\_kiyono@phys.metro-u.ac.jp} \ and Nobuko Fuchikami \\
{\small \textit{Department of Physics, Tokyo Metropolitan University, 1-1 Minami-Ohsawa, 192-0397, Japan
}}
}
\date{(\today)}
\begin{document}

\sloppy
\maketitle

\begin{abstract}
We carried out a fluid dynamical simulation for a forced dripping faucet system using a new algorithm that was recently developed. The simulation shows that periodic external forcing induces transitions from chaotic to periodic motion and vice versa. 
We further constructed an improved mass-sprig model for the same system on the basis of data obtained from the fluid dynamical computations. 
A detailed analysis of this simple model demonstrated that the periodic motion is realized after a homoclinic bifurcation, although a stable periodic orbit is generated via a Hopf bifurcation which occurs just after a saddle node one. 
\end{abstract}

\baselineskip 6.5mm

Chaos occurs widely in engineering and natural systems. In the past few years practical implementations of controlling chaos have been studied with great interest [1-6]. 
Such control schemes can be divided broadly into two categories, {\it feedback} control and {\it nonfeedback} control. The pioneering work on controlling chaos by Ott, Grebogi, and Yorke (OGY method)\cite{ogy90} is a representative example of feedback control systems. The OGY method aims to stabilize one of many unstable periodic orbits embedded in the chaotic attractor through small time-dependent perturbations in an accessible system parameter. Although the OGY method is very general, it is difficult for some high-speed systems to implement the control procedure. 
In the nonfeedback control systems, on the other hand, the applied perturbation is independent of the system's state. So far, controlling chaos by applying a suitable weak periodic perturbation, which is sometimes called {\it taming chaos} \cite{Braiman91}, has been studied in many chaotic dynamical systems (\cite{Tamura99,Mirus99} and references therein). Recently, analyzing a constrained system in which a one-dimensional Poincar\'e map is derived, Tamura {\it et al.} \cite{Tamura99} showed that taming chaos occurs by a saddle node bifurcation. However, many issues in taming chaos are still not well understood, such as phase effect \cite{Qu95}, phase diagram structures \cite{Mirus99}, and bifurcation structures for higher-dimensional dynamical systems \cite{Mirus99}. 

In this Letter, we investigate taming chaos for a dripping faucet system. The dripping faucet without forcing has been studied intensively ([7-12] and references therein) since Shaw proposed a mass-spring model \cite{shaw84}. Here we show from numerical simulations that periodic external forcing induces transitions between chaotic and periodic motion in this system. We further point out that a {\it global} homoclinic bifurcation is essential to realize the transition from chaotic to periodic motion. 

\begin{figure}[htpb]
	\begin{center}
	\includegraphics[width=0.6\linewidth]{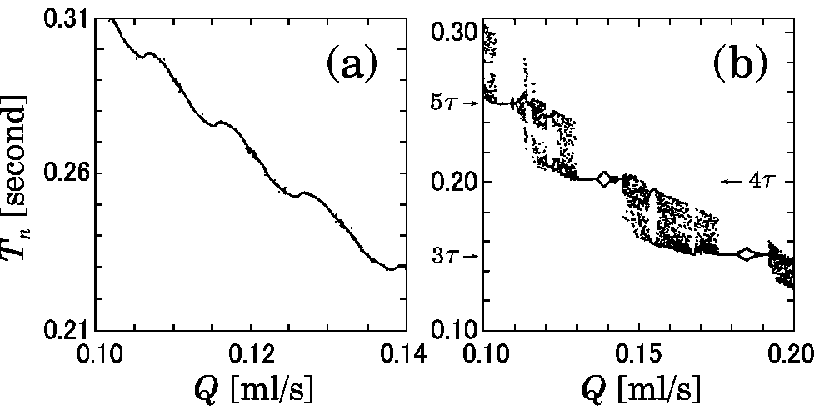}
	\end{center}
	\caption{Bifurcation diagrams in a dripping faucet experiment (a) without and (b) with periodic forcing. The faucet is 1.0 mm in inner diameter, 1.6 mm in outer diameter. In (b) the forcing period $\tau = 0.05$ s. [Reprinted by courtesy of Professor M. Shoji [12].] }
    \label{fig:2}
\end{figure}

Before going to taming chaos, we present a counter phenomenon: chaos induced by a periodic perturbation, which was observed experimentally by Shoji \cite{shoji}. 
By setting a periodically oscillating speaker on the top of the faucet, dripping time intervals $\{T_n \}$ between successive drops were measured. For sufficiently slow flow rates $Q$ without any periodic perturbation the time series $\{T_n\}$ are almost equal at each flow rate (Fig.\ \ref{fig:2} (a)). In contrast, the time series $\{T_n\}$ are distributed over a finite range for various flow rates when the speaker oscillates periodically (Fig.\ \ref{fig:2} (b)). Moreover, frequency entrainments are observed over a certain range of the flow rate. 

\begin{figure}[htpb]
	\begin{center}
	\includegraphics[width=0.6\linewidth]{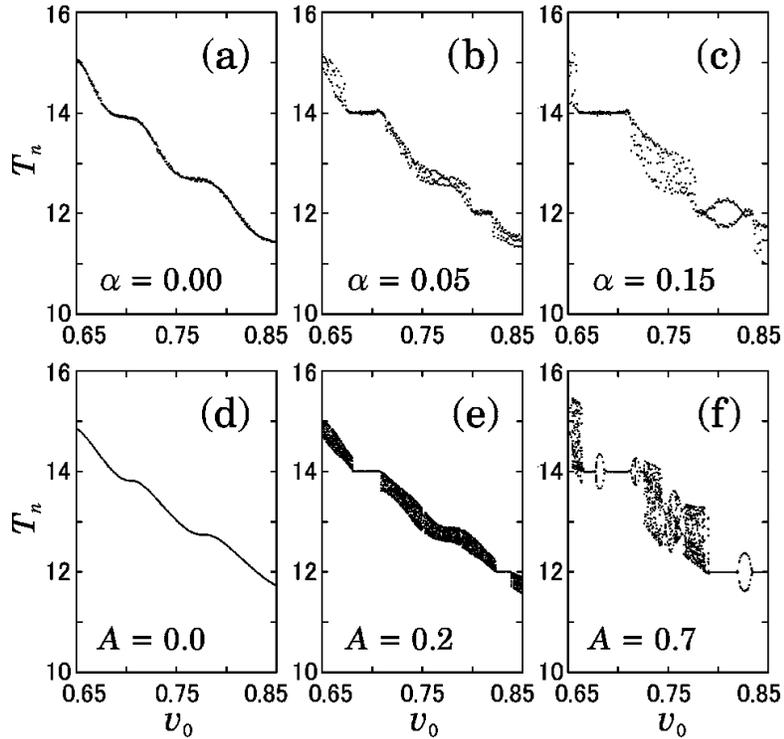}
	\end{center}
	\caption{Bifurcation diagrams (a)-(c) obtained from the fluid dynamical computations and (d)-(f) from the improved mass-spring model. (a)(d) the unperturbed system. The faucet radius $a = 0.183$ corresponds to 1.0 mm in inner diameter. $\tau = 2$ (= 0.034 s). }
    \label{fig:3}
\end{figure}

Similar results were obtained from our fluid dynamical computations (FDC) as shown in Fig.\ \ref{fig:3} (a)-(c). The algorithm of the FDC \cite{fuchi99}, which is based on Lagrangian description, was recently developed and succeeded in reproducing various complex behaviors observed experimentally in the dripping faucet under a constant flow velocity $v_0$. In the present simulation, the flow velocity oscillates as 
\begin{equation}
v = v_0 \Bigg(1 + \alpha \sin\left(\frac{2 \pi t}{\tau}\right) \Bigg) . \label{eq:force}
\end{equation}
(The flow rate is defined as $Q = \pi a^2 v_0$, where $a$ is the faucet radius). Throughout the paper the units of the length, time, and mass are chosen as $l_0 \equiv \sqrt{\Gamma / \rho g} \ (= 0.27\ {\rm cm})\,, \quad
t_0 \equiv (\Gamma/ \rho g^3)^{1/4} \ (= 0.017\ {\rm s}) \,, \quad$ and 
$m_0 \equiv \rho l_0^3 \ (= 0.020\ {\rm g}) \,, \quad
\label{eq:unit} 
$ respectively, where $\Gamma$ is the surface tension, $\rho$ is the density, and $g$ is the gravitational acceleration. (The numbers in parenthesis correspond to the water at 20~$^{\circ}$C). The viscosity was $ \eta = 0.002$ (for the water) in units of $\eta_0 \equiv (\rho \Gamma^3/g)^{1/4}$. The FDC results (Fig.\ \ref{fig:3} (a)-(c)) qualitatively agree well with the experiment, although the number of the data points are not enough because of long computational time. 

Corresponding to the above FDC, we constructed an improved mass-spring model (see \cite{kiyono99} for details) with a periodic external force: 
\begin{eqnarray}
 \frac{{\rm d}}{{\rm d}t} \left( m \frac{{\rm d}z}{{\rm d}t} \right) &=& - k z - \gamma \frac{{\rm d}z}{{\rm d}t} + m g + A \sin \left( \frac{2 \pi t}{\tau} \right),  \label{eq:ms1} \\
\frac{{\rm d}m}{{\rm d}t} &=& \pi a^2 v_0 = {\rm const.}, \label{eq:massi}
\end{eqnarray}
where $z$ is the center of mass of the forming drop, and $m$ is its mass. 
The damping parameter $\gamma$ was chosen as $\gamma = 0.008$. 
The spring constant $k$ depends on the mass as
\begin{equation}
k\left( m \right) = 
\left\{ 
	\begin{array}{cl}
	 3.1 m^{-0.58} & ( m < 0.584 ) \\
	 -31 m^2 + 32 m - 3.88 & ( 0.584 \le m < 0.891 ) \\
	 0 & ( m \ge 0.891 ) 
	
	\end{array}
	\label{eq:kfunk1}
\right. .
\end{equation}
Breakup of a drop is taken into account by assuming that a part of the mass is lost when the position $z$ reaches a critical point $z_{{\rm crit}}$. The remnant mass under the faucet is renewed as
\begin{equation}
m_{\rm r} = B m - C , \hspace{.5cm} {\rm when} \hspace{.5cm} z = z_{{\rm crit}},
\label{eq:reset1}
\end{equation}
depending on the total mass $m$ at the breakup moment, where $B = 0.068$ and $C = 0.053$. The position and velocity just after a breakup moment are renewed as
\begin{equation}
\left.
\begin{array}{lll}
z &=& z_0  \\
\dot{z} &=& 0 \\
\end{array}
\right\} \hspace{.5cm} {\rm when} \hspace{.5cm} z = z_{{\rm crit}}. 
\label{eq:reset2}
\end{equation}
The breakup position $z_{{\rm crit}}$ and the renewed position $z_0$ were assumed as constant: $z_0 = 0.15$ and $z_{{\rm crit}} = 4$. The relations (\ref{eq:kfunk1}), (\ref{eq:reset2}) and the above parameters were obtained by referring to the results of the FDC for $a = 0.183$. 

When no external force is applied ($A = 0$), the remnant mass at the breakup moment, $m_{{\rm r},n+1}$, is uniquely determined only by the previous remnant mass $m_{{\rm r},n}$, i.e., Poincar\'e map is defined as $m_{{\rm r},n+1} = M(m_{{\rm r},n})$, even though the return map of $T_n$ is not necessarily single-valued. In fact, This significant feature of the dripping faucet system was first found from the FDC \cite{kiyono99}. Under external forcing, on the other hand, the Poincar\'e map just after the breakup moment is two dimensional because of an additional degree of freedom, the phase of the external force. The Poincar\'e map is thus expressed as
\begin{equation}
\left( \begin{array}{c}
m_{{\rm r}, n+1} \\
\theta_{n+1} \\
\end{array} \right) = {\bf M} \left( \begin{array}{c}
								m_{{\rm r}, n} \\
								\theta_{n} \\
								\end{array} \right) \, , \label{2dmap}
\end{equation}
where
\[ 
\theta = \frac{t}{\tau}\ {\rm modulo}\ 1.
\]

As shown in Fig.\ \ref{fig:3} (d)-(f), the mass-spring model can reproduce a bifurcation diagram in good agreement with the FDC (Fig.\ \ref{fig:3} (a)-(c)). Moreover, the agreement with the experiment (Fig.\ \ref{fig:2} (a), (b)) is even better than for the FDC, because the numerical computations of this simple model can be carried out until transient behavior has completely vanished.

\begin{figure}[htpb]
	\begin{center}
	\includegraphics[width=0.6\linewidth]{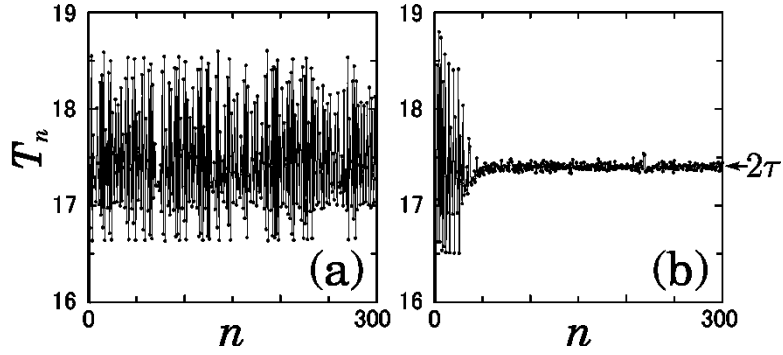}
	\end{center}
	\caption{Time series of $T_n$ obtained from the fluid dynamical computations for (a) unperturbed and (b) periodically perturbed ($\alpha = 0.05$, $\tau = 8.7$) system. $a = 0.916$, $v_0 = 0.95$, $N = 2$. }
    \label{fig:4}
\end{figure}

Contrary to the above phenomenon, it is also possible to eliminate chaotic motion by applying a suitable periodic perturbation (taming chaos). An experiment by Katsuyama and Nagata for the faucet of about 5 mm in diameter, in which no periodic perturbation was applied, showed that a unit structure including chaotic transitions repeatedly appears in a bifurcation diagram in a wide range of the flow rate \cite{kn98}. A similar repeating structure in the bifurcation diagram was obtained from both our FDC \cite{fuchi99} and our improved mass-spring model \cite{kiyono99}. 
Figure \ref{fig:4}(a) is an example of a chaotic oscillation of $T_n$ obtained from the FDC for the faucet radius $a = 0.916$ (2.5 mm) and the fluid velocity $v_0 = 0.95$, where no perturbation is applied. This chaotic oscillation is eliminated by a periodic perturbation (Eq.\ (\ref{eq:force}) with $\alpha = 0.05$ and $\tau = 8.7$) and a period-one (P1) motion with a period $2\tau$ is generated (Fig.\ \ref{fig:4}(b)). 

\begin{figure}[htpb]
	\begin{center}
	\includegraphics[width=0.6\linewidth]{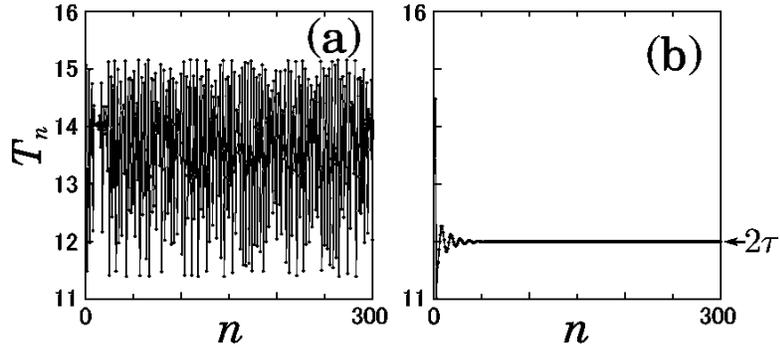}
	\end{center}
	\caption{Time series of $T_n$ obtained from the improved mass-spring model for (a) unperturbed and (b) periodically perturbed ($A = 1.2$, $\tau = 6$) system. $a = 0.916$, $v_0 = 0.115$, $N = 2$. }
    \label{fig:6}
\end{figure}

In the mass-spring model, a P1 motion (Fig.\ \ref{fig:6}(b)) is also induced from a chaotic state (Fig.\ \ref{fig:6}(a)) by periodic forcing. 
Referring to data obtained from the FDC for $a=0.916$ (0.25 mm), parameters of the mass-spring model were chosen as ($\gamma = 0.05$, $z_{{\rm crit}}=5.5$, $z_0 = 2.0$, $B=0.2$, and $C=0.3$), and the spring constant was assumed as 
\begin{equation}
k\left( m \right) =
\left\{ 
	\begin{array}{cl}
	 -11.4 m + 52.5 & ( m < 4.61 ) \\
	 0 & ( m \ge 4.61 ) 
	
	\end{array}
	\label{eq:kfunk5}
\right. . \\
\end{equation}
Figure \ref{fig:6} shows that, for the forcing amplitude $A = 1.2$ and the period $\tau = 6$, the system is entrained to the P1 motion with the period $2 \tau$. 

\begin{figure}[htpb]
	\begin{center}
	\includegraphics[width=0.6\linewidth]{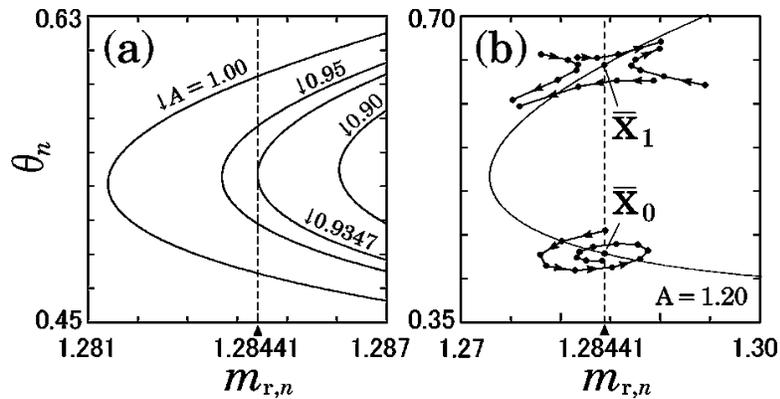}
	\end{center}
	\caption{(a) $I$ is shown as a dashed line and their images ${\bf M}(I)$ are shown as solid curves at various forcing amplitudes $A$. (b) Orbits near period-one orbits on the Poincar\'e section. }
    \label{fig:7}
\end{figure}

\begin{figure}[htpb]
	\begin{center}
	\includegraphics[width=0.6\linewidth]{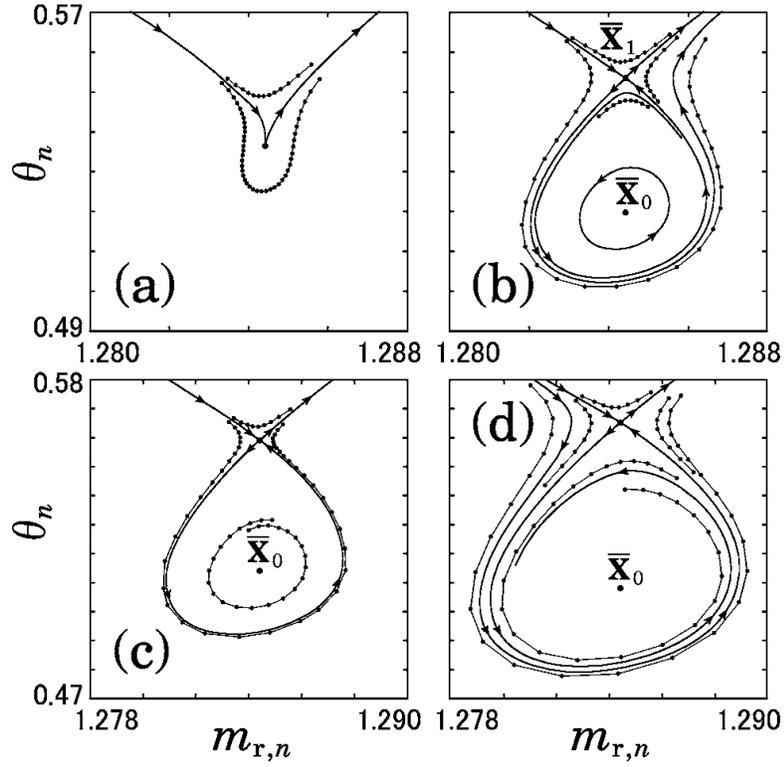}
	\end{center}
	\caption{Invariant manifolds (solid curves with arrows) and examples of orbits (dots and thin lines). $v_0 = 0.115$, $\tau=6$. (a) $A = 0.9347$. (b) $A = 0.940$. (c) $A = 0.9441$. (d) $A = 0.950$. }
    \label{fig:8}
\end{figure}

\begin{figure}[htpb]
	\begin{center}
	\includegraphics[width=0.6\linewidth]{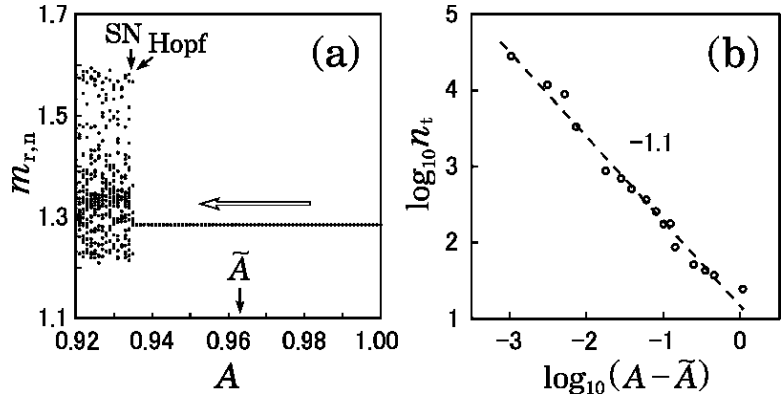}
	\end{center}
	\caption{(a) Log-log plot of the average transient lifetime $n_{\rm t}$ vs $(A-\tilde{A})$, where $\tilde{A} = 0.963$. (b) Bifurcation diagram of $T_n$ with decreasing $A$. }
    \label{fig:9}
\end{figure}

 We shall investigate how the entrainment is realized in the dripping faucet by analyzing the Poincar\'e section of the mass-spring model in detail. 
Let ${\bf x} \equiv \left( m_{{\rm r}, n}, \theta_n \right)$ be a state point on the Poincar\'e section just after the breakup moment (Eq.\ (\ref{2dmap})), and let $\overline{\bf x} \equiv \left( \overline{m}_{{\rm r}}, \overline{\theta} \right)$ be a P1 orbit satisfying $\overline{\bf x} = {\bf M} ( \overline{\bf x} )$. If the P1 orbit whose period is an integer multiple of the forcing period: $T_n = N \tau = {\rm const.}$, then the linear relation Eq.\ (\ref{eq:reset1}) yields an explicit expression for $\overline{m}_{{\rm r}}$: 
\begin{equation}
\overline{m}_{\rm r} = \frac{BQN\tau+C}{1-B} \hspace{0.3cm} ,\ Q = \pi a^2 v_0 . \label{mr5}
\end{equation}
The P1 orbits $\overline{\bf x}$ should then satisfy
\begin{equation}
\overline{\bf x} \in I \cap {\bf M}(I), \label{po1}
\end{equation}
where $I \equiv \{ (m_{\rm r}, \theta) \mid m_{\rm r} = \overline{m}_{\rm r} \}$, and the entire line $I$ is mapped onto a curve ${\bf M}(I)$. 
For a specified value of the integer $N$, the P1 orbits are obtained from Eq.\ (\ref{mr5}). 
Figure \ref{fig:7}(a) illustrates how the P1 orbits with $N = 2$ are generated as the forcing amplitude $A$ is increased, where $I$ is shown by a dashed line and ${\bf M}(I)$ is shown by a solid curve. As $A$ grows through $A=0.9347$, $I$ intersects ${\bf M}(I)$. It was found numerically that two P1 orbits $\overline{\bf x}_0$ and $\overline{\bf x}_1$, i.e., the intersection of $I$ and ${\bf M}(I)$, are generated via a saddle node bifurcation where $\overline{\bf x}_1$ is a saddle. 
Despite the terminology `saddle node', however, the other fixed point $\overline{\bf x}_0$ is a source instead of a sink just after the saddle node bifurcation point. Therefore, there is only one attractor which is chaotic at this stage. It was found from a detailed analysis that bifurcations of the Poincar\'e map of the present system can well be explained by a discretized version of a 2-dimensional flow which yields a homoclinic bifurcation: A scenario to a homoclinic bifurcation similar to that presented in Ref.\ \cite{wiggins} can be applied. Figure \ref{fig:8} illustrates invariant manifolds of the Poincar\'e map (solid curves with arrows) together with numerical examples of orbits near the manifolds (dots and thin lines). The saddle node bifurcation point is presented in Fig.\ \ref{fig:8}(a). 
As the parameter $A$ increases, the unstable fixed point $\overline{\bf x}_0$ is stabilized by a Hopf bifurcation (Fig.\ \ref{fig:8}(b)). The basin boundary of $\overline{\bf x}_0$ is then an unstable limit cycle generated by the Hopf bifurcation. 
Further increase of $A$ induces a homoclinic bifurcation (Fig.\ \ref{fig:8}(c)). Up to this stage, the basin of $\overline{\bf x}_0$ is a small closed region. After the homoclinic bifurcation, the basin of $\overline{\bf x}_0$, whose boundary is a heteroclinic orbit, is open. However, the chaotic attractor coexists and the motion remains chaotic because the basin of $\overline{\bf x}_0$ is still small. After $A$ passes through the critical value $\tilde{A} = 0.963$, the chaotic attractor disappears and the periodic motion $\overline{\bf x}_0$ is realized. 
The coexistence of the two attractors results in a hysteresis as shown in Fig.\ \ref{fig:9}(a). If one starts from a large enough value of $A$ (say $A = 1.2$, Fig.\ \ref{fig:7}(b)) at which the P1 motion is realized, and decreases the $A$ value through $\tilde{A}$, the system keeps the P1 motion till the Hopf bifurcation occurs at $A \approx 0.936$. 
At the critical point $A = \tilde{A}$, the system exhibits the following scaling property: 

\begin{equation}
n_{\rm t} \sim ( A - \tilde{A} )^{-1.1} \hspace{0.4cm} {\rm for} \  A>\tilde{A} \, . \label{scaling}
\end{equation}
Here $n_{\rm t}$ is the average transient lifetime defined by the average iteration number before the orbit falls within a distance $\Delta$ from the stable P1 orbit; the average was taken over many initial points on the Poincar\'e section. The exponents and the value of $\tilde{A}$ were determined by least square fitting. 

It is interesting that a similar structure of the Poincar\'e section as in Fig.\ \ref{fig:8}(d) was observed in a dripping faucet experiment by Pinto and Sartorelli \cite{pinto00}, although no periodic force was applied. The present analysis suggests that their result may possibly be explained if any oscillatory factor was generated inevitably in the experimental condition.

In conclusion, we have demonstrated that periodic forcing induces transitions between chaotic and periodic motions in the realizable dripping faucet system. The dimension of the dynamical system generally increases by 1.0 when a forced perturbation is added. Then, a saddle node bifurcation that is not inherent in the original dynamics may occur. 
Higher-dimensional systems such as the dripping faucet may induce a global bifurcation further. 
It was shown that the change of the global structure in the phase space is essential in realizing the entrainment for the present system. 
Postnov {\it et al.} \cite{postnov99} analyzed a periodically forced 4-dimensional flow, and found that a torus coexists with a stable periodic orbit between a saddle node bifurcation and a homoclinic one. In the present case, in contrast, chaos coexists with a stable periodic orbit between a Hopf bifurcation and a homoclinic one. 

We thank Professor M. Shoji for permission to copy his figure. This research was partly supported by Iketani Science and Technology Foundation.




\end{document}